\documentstyle[11pt,paspconf,psfig]{article}

\begin{document}

\title{The CFH Optical PDCS Survey (COP): First results}
\author{C.Adami, B.Holden, A.Mazure, F.Castander, R.Nichol, M.Ulmer, 
M.Postman, L.Lubin}
\affil{AA: IGRAP-LAS (Marseille) / Northwestern University (Chicago), 
AB: University of Chicago, AC: IGRAP-LAS, 
AD: Observatoire Midi-Pyr. (Toulouse) / University of Chicago,
AE: Carnegie Mellon University (Pittsburgh), 
AF: Northwestern University, AG: STScI, AH: Cal. Tech.}

\begin{abstract}
We present in this paper the first results of the COP survey about the
reality of the PDCS clusters, about their velocity dispersions and dynamic and
about the periodicity of the structures along the line of sight.
\end{abstract}


\section{The data}

\subsection{Description of the survey}

We have made a spectroscopical follow-up of 10 PDCS lines of sight including
15 clusters. These PDCS clusters (Lubin et al. 1996) are optically selected
cluster candidates. We have measured about 700 redshifts in 6 nights at CFH 
with the MOS spectrograph. The expected redshift of these clusters is 
around 0.4. Two areas of the sky are particularly well covered: 4 lines of 
sight around 9h and 3 lines of sight around 13h. The sampled redshift 
range is z=[0.,0.9].

\subsection{Are PDCS clusters real?}

The first results of this survey show that more than 60\% of the candidate
clusters are real gravitationally bounded structures with a velocity 
dispersion ranging from 600 km/s to 1500 km/s. 

\section{Results}

\subsection{Dynamic of the clusters}

Using the 6 best sampled clusters, we have used the same techniques as were
were used for the nearby ENACS clusters (Adami et al. 1998). We plot the 
variation of the galaxy normalized velocity dispersion versus the absolute 
magnitude and the morphological type (fig 1: we assume as a first approximation 
that the em line galaxies are a mix of early and late spirals and that the only 
absorption line galaxies are a mix of ellipticals and SO). The open symbols are 
the results for nearby clusters (z$\simeq$0.05: Adami et al. 1998) and the 
filled triangles are for the PDCS clusters (z$\simeq$0.4). We see that the two
distributions are essentially the same: the evolution of the internal dynamics
of the clusters between 0.05 and 0.4 seems to be negligible. At z$\simeq$0.4 or 
z$\simeq$0.05, the emission line galaxies are an infalling population while the 
absorption line galaxies seem to be virialized and the galaxies follow the 
energy equipartition law (thick line in fig 1 left). The conclusion is that the 
epoch formation of the clusters is probably significantly greater than 0.4, the
clusters continually evolving after with late type galaxies still infalling at
low redshifts.

\begin{figure} 
\vbox 
{\psfig{file=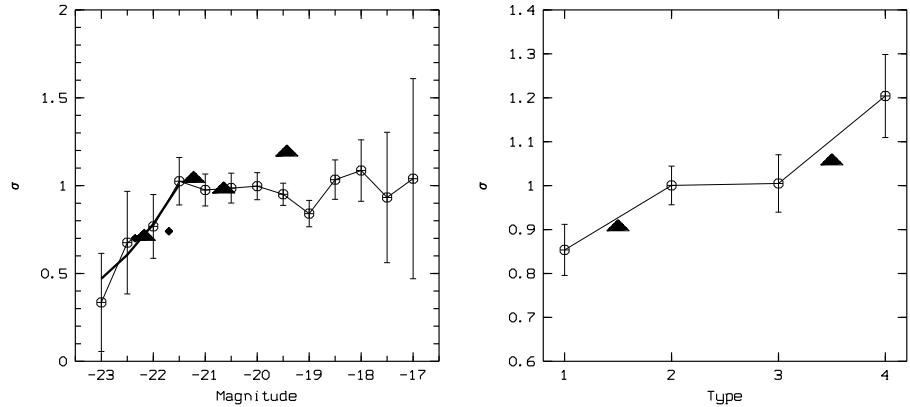,width=13.cm,angle=270}} 
\caption[]{(left panel/Open symbols): normalized velocity dispersion vs. R 
magnitude for the nearby cluster galaxies. We show the relation 
$\sigma$=10$^{0.2 mag.}$, normalized at R=-21.5 ; (right panel/Open symbols): 
normalized velocity dispersion vs. morphological type for the nearby cluster
galaxies with 1: Ellipticals, 2: SO, 3: Early spirals, 4: Late Spirals ;
(left and right panel / large filled symbols): z$\simeq$0.4 cluster galaxies.
(left panel / small filled symbols): results when we split the brighter bin in
two bins of 8 galaxies.} 
\label{} 
\end{figure}

\subsection{Periodicity of the structures in the COP survey}

The 2 well sampled COP areas (9h and 13h) allow a study of the periodicity
along the line of sight. The structures are defined exactly as the 
ENACS (Katgert et al. 1996). At 9h, we find a periodicity of 90 
$\pm$ 2 Mpc and at 13h the periodicity is 143 $\pm$ 10 Mpc. Comparing these
results with Broadhurst et al (1990) (128 Mpc in another direction), we
tentatively conclude that periodicities in the structure distribution are in 
agreement with the "Web" representation of the Universe and the value
of this periodicity depends of the line of sight.

\end{document}